\documentclass[aps,12pt,pra,showpacs,superscriptaddress,preprint,a4paper]{revtex4}
\usepackage{amsmath}
\usepackage{subfig}
\usepackage{array}
\usepackage{amsfonts}
\usepackage{epsf}
\usepackage{epsfig}
\usepackage{amssymb}
\usepackage{bbm}
\usepackage{delarray}
\usepackage{xcolor}
\catcode`ð=\active
\defð{\u{g}}
\catcode`Ð=\active
\defÐ{\u{G}}
\catcode`Ý=\active
\defÝ{\. I}
\catcode`ö=\active
\defö{\"{o}}
\catcode`Ö=\active
\defÖ{\"O}
\catcode`ü=\active
\defü{\"{u}}
\catcode`Ü=\active
\defÜ{\"{U}}
\catcode`Þ=\active
\defÞ{\c{S}}
\catcode`þ=\active
\defþ{\c{s}}
\catcode`ý=\active
\defý{{\i}}
\catcode`ç=\active
\defç{\d{c}}
\catcode`Ç=\active
\defÇ{\d{C}}

\begin{document}
\title{Klein-Gordon equation for a charged particle in space varying electromagnetic fields-A systematic study via Laplace transform }
\author{\small Tapas Das}
\email[E-mail: ]{tapasd20@gmail.com, arda@hacettepe.edu.tr}\affiliation{Kodalia Prasanna Banga High
 School (H.S.), South 24 Parganas, 700146, India}
\author{\small Altuð Arda}
\altaffiliation[Present adress: ]{Department of Mathematical Science, City University London,
Northampton Square,\\ London EC1V 0HB, UK}\affiliation{Department of
Physics Education, Hacettepe University, 06800, Ankara,Turkey}
\begin{abstract}
Exact solutions of the Klein-Gordon equation for a charged particle in the presence of three spatially varying electromagnetic fields, namely, (i) $\vec{E}=\alpha\beta_0e^{-\alpha x_2}\hat{x}_2$,  $\vec{B}=\alpha\beta_1e^{-\alpha x_2}\hat{x}_3$ (ii) $\vec{E}=\frac{\beta_0^{'}}{x_2^2}\hat{x}_2$, $\vec{B}=\frac{\beta_1^{'}}{x_2^2}\hat{x}_3$, and (iii) $\vec{E}=\frac{2\beta_0^{'}}{x_2^3}\hat{x}_2$, $\vec{B}=\frac{2\beta_1^{'}}{x_2^3}\hat{x}_3$, are studied. All these fields are generated from a systematic study of a particular type of differential equation whose coefficients are linear  in independent variable. The Laplace transform approach is used to find the solutions and the corresponding eigenfunctions are expressed in terms of the hypergeometric functions $\,_{1}F_{1}(a', b'; x)$ for first two cases of the above configurations while the same are expressed in terms of the Bessel functions of first kind, $J_{n}(x)$, for the last case.

\end{abstract}
\maketitle
\newpage
\section{Introduction}
Proper use of the non-relativistic wave equation, i.e, Schrödinger equation, still helps to understand the deep phenomenological aspects of the different subject matters generally, covered by nuclear physics, molecular physics and quantum chemistry [1-3]. However, when the particle travels very close to the speed of light, we have to take the relativistic wave equations such as Klein-Gordon (KG) or Dirac equation. Like the Schrödinger equation the KG equation also describes a spin zero particle. This suggests that these two equations are very similar in nature except the situation whether it is non-relativistic or relativistic. Based on the above phenomenological facts there has been an increasing interest in the study of relativistic wave equations in recent years, particularly KG equation [4-10].

After the profound success of Maxwell's equations for classical electrodynamic, quantum mechanics provides a wide space to study the physics of different events very close to the practical purpose. Nowadays, there has been a tendency to analyse the wave equations, whether it is relativistic or non relativistic, for a charged particle moving in external electromagnetic (em) field [11-13]. As an application, em fields with different spatial configurations have been used in a wide range of physics, from the high energy physics (within the Standard Model as a relativistic quantized field theory, for example) [14] to the biophysics relating with the effect of em fields on cells [15]. These type of studies, whether relativistic or non relativistic, always indicate how the model of quantum mechanical systems under external em fields able to describe the different physical events very near to the practical implication. Such systems have been studied by many authors focusing especially on exact solutions [16, 17 and references therein]. Moreover applied optics is one of the vast area to study relativistic quantum mechanics. In the strong laser field when the refractive index of the material is less than unity, exact solutions of Klein-Gordon equation helps to understand the interaction of the charged particles with the external em field [18]. The reference list [19-22] covers the many other useful uses of the quantum mechanical wave equations and also different em field profiles that are used, starting from electrical engineering to magneto-plasma physics.

The effect of an electromagnetic field on a charged particle studied lot and in spite of its long history [23-26],
still requires additional study in terms of different mathematical methods for various em field profiles. Motivated by these circumstances, in this paper we try to examine exact solutions of the KG equation for some different spatially-dependent em profiles by the means of the Laplace transformation method which is a very elegant technique, and can be used to transform a second order linear differential equation with coefficients that are linear in independent variable into a first order one [27-30]. Various useful properties of this integral transform ease out the scenario of finding energy eigenvalues and corresponding eigenfunctions.

The present paper is arranged as follows. Section II is a brief outline of the Laplace transformation method. In Section III, we study the model differential equation originated from KG equation and make it compatible for the Laplace transform approach. This study generates a few spatially-dependent, time-independet electromagnetic fields. In section IV, we solve the KG equation for different electromagnetic field configurations, where electric and magnetic fields are orthogonal, for obtaining the bound states and the corresponding eigenfunctions. Conclusion of the work is placed in Section V.

\section {Overview of Mathematical Methods}
The Laplace transform $\chi(y)$ or $\mathcal{L}$ of a function $f(t)$ is defined by [31]
\begin{eqnarray}
\chi(y)=\mathcal{L}\left\{{f(t)}\right\}=\int_{0}^{\infty}e^{-{yt}}{f(t)}dt\,.
\end{eqnarray}
If there is some constant $\sigma\in \Re$ such that ${\left|e^{-{\sigma}{t}}{f(t)}\right|\leq M}$
for sufficiently large $t$, the integral in Eq. (1) will exist for  Re $y>\sigma$ . The Laplace transform may fail to exist because of a sufficiently strong singularity in the function $f(t)$ as $t\rightarrow 0$ . In particular
\begin{eqnarray}
\mathcal{L}\left[\frac{t^{\nu}}{\Gamma(\nu+1)}\right]=\frac{1}{y^{\nu+1}}\,,{\nu}>-1\,.
\end{eqnarray}
The Laplace transform has the following derivative properties [32]
\begin{subequations}
\begin{align}
\mathcal{L}\left\{t^pf^{(n)}(t)\right\}|_{p\geq n}=\left(-\frac{d}{dy}\right)^p[y^n\chi(y)]\,,\\
\mathcal{L}\left\{t^pf^{(n)}(t)\right\}|_{p<n}=\left(-\frac{d}{dy}\right)^p[y^n\chi(y)]+(-1)^{p-1}\Bigg[\frac{(n-1)!}{(n-p-1)!}y^{n-p-1}f(0)+\nonumber \\ \frac{(n-2)!}{(n-p-2)!}y^{n-p-2}f^{'}(0)+\cdots+p!f^{(n-p-1)}(0)\Bigg]\,.
\end{align}
\end{subequations}
When $p=0$, it reduces to
\begin{eqnarray}
\mathcal{L}\left\{f^{(n)}(t)\right\}=y^n\chi(y)-\sum_{k=0}^{n-1}y^{n-1-k}{f^{(k)}(0)}\,,
\end{eqnarray}
while for $n=0$
\begin{eqnarray}
\mathcal{L}\left\{t^{p}f(t)\right\}=(-1)^{p}\chi^{(p)}(y)\,.
\end{eqnarray}
where the superscript such $(j)$ denotes the $j$-th derivative with respect to $t$ for $f^{(j)}{(t)}$, and with respect to $y$
for $\chi^{(j)}{(y)}$.
The inverse transform is defined as $\mathcal{L}^{-1}\left\{\chi(y)\right\}=f(t)$. One of the most important properties of the Laplace transform is that given by the convolution theorem [33]. This theorem is a powerful method to find the inverse Laplace transform. According to this theorem if we have two transformed function $g(y)=\mathcal{L}\left\{G(t)\right\}$ and
$h(y)=\mathcal{L}\left\{H(t)\right\}$, then the product of these two is the Laplace transform of the convolution $(G*H)(t)$, where
\begin{eqnarray}
(G*H)(t)=\int_{0}^t G(t-\tau)H(\tau)d\tau\,,
\end{eqnarray}
So the convolution theorem yields
\begin{eqnarray}
\mathcal{L}(G*H)(t)=g(y)h(y)\,,
\end{eqnarray}
Hence
\begin{eqnarray}
\mathcal{L}^{-1}\left\{g(y)h(y)\right\}=\int_{0}^t G(t-\tau)H(\tau)d\tau\,.
\end{eqnarray}
if we substitute $w=t-\tau$, then we find the important consequence $G*H=H*G$.

\section{Mathematical Model}
The Klein-Gordon equation in \textit{natural unit} for a particle having the rest mass $m$, and  the electrical charge $q$
is written as
\begin{eqnarray}
(\square+m^2)\Psi(\vec{\bold{x}}, t)=0\,,
\end{eqnarray}
where the operator $\square=\frac{\partial^2}{\partial t^2}-\nabla^2=\partial_\mu\partial^\mu$ is the d'Alembertian [34].
In an electromagnetic field, the above equation is written by using the ``minimal substitution''
\begin{eqnarray}
(D{_\mu}D^{\mu}+m^2)\Psi(\vec{\bold{x}}, t)=0\,,
\end{eqnarray}
where $D_{\mu}=\partial_{\mu}+iqA_{\mu}$ is covariant derivative and $A_{\mu}=\{A_{0},\vec{\bold{A}}\}$ is the four-potential of the electromagnetic field. We will consider that the components of $A_{\mu}$ is dependent on space variable only spanned by the unit vectors $\hat{x}_{i}\,(i=1, 2, 3)$. Introducing the renowned gauge condition
$\partial_\mu A^{\mu}=0$ (Lorentz gauge), we have the equation of motion
\begin{eqnarray}
\left\{\square+2iqA_{\mu}\partial^{\mu}-q^2A^{\mu}A_{\mu}+m^2\right\}\Psi(\vec{\bold{x}}, t)=0\,,
\end{eqnarray}
where we have used the common fact $A^{\mu}\partial_{\mu}=A_{\mu}\partial^{\mu}$.
The vector potential $A_{\mu}=\{A_{0},\vec{\bold{A}}\}$ can generate unique field vectors $\vec{E}$ and $\vec{B}$ via the relation $\vec{B}=\nabla\times\vec{A}$ and $\vec{E}=-\nabla A_0-\frac{\partial\vec{A}}{\partial t}$. In this paper, we will concentrate on the time-independent, space varying physical fields. In general the profile of our field has the following form $A_{\mu}=\{A_{0},\vec{\bold{A}}\}=(\beta_0s(x_2),\beta_1s(x_2),0,0)$ where $\beta_{i}$'s ($i=0, 1$) are constants, and $s(x_2)$ is the unknown function which determines the required field for the implication of practical purpose. Now considering the above field profile and taking solution as
\begin{eqnarray}
\Psi(\vec{\bold{x}}, t)=e^{i(x_1p_1+x_3p_3-Et)}\phi(x_2)\,,
\end{eqnarray}
where $p_1$ is the linear momentum of the particle along the $x_1$ axis, and $p_3$ is the same
along $x_3$ axis, Eq. (11) changes to
\begin{eqnarray}
\Big(\frac{d^2}{dx_2^2}+\lambda_0+\lambda_1s(x_2)+\lambda_2s^2(x_2)\Big)\phi(x_2)=0\,,
\end{eqnarray}
with the following abbreviations
\begin{subequations}
\begin{align}
\lambda_0&=E^2-p_1^2-p_3^2-m^2 \,,\\
\lambda_1&=2q(\beta_1p_1-E\beta_0)\,,\\
\lambda_2&=q^2(\beta_0^2-\beta_1^2)\,.
\end{align}
\end{subequations}
Now, let us change the independent variable from $x_2$ to $t=t(x_2)$ and set $\phi(x_2)=t^\gamma f(t)$. After few algebraic steps, we arrive at
\begin{eqnarray}
\Big(\frac{dt}{dx_2}\Big)^2f^{''}&+&\Big[2\gamma\frac{1}{t}\Big(\frac{dt}{dx_2}\Big)^2+\frac{d^2t}{dx_2^2}\Big]f^{'}\nonumber\\&+& \Big[\gamma(\gamma-1)\frac{1}{t^2}\Big(\frac{dt}{dx_2}\Big)^2+\frac{\gamma}{t}\frac{d^2t}{dx_2^2}+\lambda_0+\lambda_1s(x_2)+\lambda_2s^2(x_2)\Big]f=0\,,
\end{eqnarray}
where $f^{'}=\frac{df}{dt}$ and $f^{''}=\frac{d^2f}{dt^2}$. In order to deduce a potentially second order differential equation with coefficients linear in $t$, we  make further choices
\begin{eqnarray}
\frac{dt}{dx_2}=\alpha t^{m_1} \,,\, s(x_2)=t^{m_2}\,,
\end{eqnarray}
where $m_1,m_2$ and $\alpha$ are constants. Consequently, the resultant equation reads (with $\widetilde{\lambda}_i=\frac{\lambda_i}{\alpha^2}$)
\begin{eqnarray}
t^{2m_1}f^{''}+(2\gamma+m_1)t^{2m_1-1}f^{'}+[\gamma(\gamma+m_1-1)t^{2m_1-2}+\widetilde{\lambda}_0+\widetilde{\lambda}_{1}t^{m_2}+\widetilde{\lambda}_{2}t^{2m_2}]f=0\,.
\end{eqnarray}
The interesting fact of the above equation is that it is independent of the sign of the constant
$\alpha$. So, to get physically and practically accepted vector potentials we are free to choose the sign of $\alpha$ without hampering the main mathematics. There are several possibilities that can transform the above equation into a second order differential equation with coefficients linear in $t$. We exemplify some of them
\begin{enumerate}
\item{Case-I}\\
If we choose $\gamma(\gamma+m_1-1)t^{2m_1-2}+\widetilde{\lambda}_0=0$ , it implies $m_1=1$ and
$\widetilde{\lambda}_0=-\gamma(\gamma+m_1-1)$. For this, selecting $\alpha\rightarrow-\alpha$ we have $t^{'}=\frac{dt}{dx_2}=-\alpha t$ and $t=e^{-\alpha x_2}$. The remaining equation can be made linear coefficients in $t$ by setting $m_2=1$, which results in
\begin{eqnarray}
tf^{''}+(2\gamma+1)f^{'}+[\widetilde{\lambda}_1+\widetilde{\lambda}_2 t]f=0\,.
\end{eqnarray}
\item{Case-II}\\
If we take $\gamma(\gamma+m_1-1)t^{2m_1-2}+\widetilde{\lambda}_{2}t^{2m_2}=0$, it implies $m_2=m_1-1$ and $\widetilde{\lambda}_2=-\gamma(\gamma+m_1-1)$. The equation with linear coefficients is reached by setting $m_1=0$ or $m_2=-1$ as
\begin{eqnarray}
tf^{''}+2\gamma f^{'}+[\widetilde{\lambda}_{0}t+\widetilde{\lambda}_1]f=0\,.
\end{eqnarray}
Eventually, we have $t^{'}=\frac{dt}{dx_2}=\alpha $ and $t=\alpha x_2$ , where $\alpha>0$.
\item{Case-III}\\
If we now set $\gamma(\gamma+m_1-1)t^{2m_1-2}+\widetilde{\lambda}_{1}t^{m_2}=0$, it implies $m_2=2(m_1-1)$ and $\widetilde{\lambda}_1=-\gamma(\gamma+m_1-1)$. When $\widetilde{\lambda}_0=0$, two possible equations with linear coefficients can be deduced:

(a) for $m_1=2$, so that $t^{'}=\alpha t^2$ and $t=\frac{1}{\alpha x_2}\,,\alpha<0$

(b) for $m_1=\frac{3}{2}$,  so that $t^{'}=\alpha t^{\frac{3}{2}}$ and $t=\frac{4}{\alpha^2 x_2^2}\,,\alpha<0$

and the corresponding equations are
\begin{subequations}
\begin{align}
tf^{''}+2(\gamma+1)f^{'}+\widetilde{\lambda}_{2}tf&=0\,,\\
tf^{''}+2(\gamma+\frac{3}{4})f^{'}+\widetilde{\lambda}_{2}f&=0\,.
\end{align}
\end{subequations}
\item{Case-IV}\\
If we set $\gamma(\gamma+m_1-1)t^{2m_1-2}+\widetilde{\lambda}_{1}t^{m_2}=0$ but $\widetilde{\lambda}_2=0$, it implies again that $m_2=2(m_1-1)$ and $\widetilde{\lambda}_1=-\gamma(\gamma+m_1-1)$. We obtain two equations with linear coefficients:

(a) for $m_1=0$ implying $t^{'}=\alpha $ and $t=\alpha x_2$\,,

(b) for $m_1=\frac{1}{2}$ implying $t^{'}=\alpha t^{\frac{1}{2}} $ and $t=\frac{1}{4}\alpha^2 x_2^2$\,,

and the corresponding equations are
\begin{subequations}
\begin{align}
tf^{''}+2\gamma f^{'}+\widetilde{\lambda}_{0}tf&=0\,,\\
tf^{''}+2(\gamma+\frac{1}{4})f^{'}+\widetilde{\lambda}_{0}f&=0\,.
\end{align}
\end{subequations}
\end{enumerate}

\section{Bound state spectrum for different field profiles}
In this section, firstly, we will use the model equations of previous section one by one to realize different electromagnetic field profiles. Secondly, our focus will be to provide the bound state spectrum of the KG particle in those electromagnetic fields via Laplace transform method.

\subsection{Case-I}

As we derived in this case $s(x_2)=t=e^{-\alpha x_2}$, the four vector potential can be given by
$\{A_{0},\vec{\bold{A}}\}=\left\{\beta_{0}e^{-\alpha x_2}, \beta_1e^{-\alpha x_2}, 0, 0\right\}$ which gives electric and magnetic fields as $\vec{E}=\alpha\beta_0e^{-\alpha x_2}\hat{x}_2$, and $\vec{B}=\alpha\beta_1e^{-\alpha x_2}\hat{x}_3$, respectively. In this configuration, exponentially varying electric and magnetic fields are orthogonal to each other, where  $\beta_{0}$, $\beta_{1}$ are constants and $\alpha$ is a constant parameter controlling the strength of the external fields.

Now selecting $\widetilde{\lambda}_1=-\gamma_1^2$ and $\widetilde{\lambda}_2=-\gamma_2^2$ Eq. (18) can be written as
\begin{eqnarray}
\Big[t\frac{d^2}{dt^2}-(2\gamma_0-1)\frac{d}{dt}-\gamma_1^2-\gamma_2^2 t\Big]f=0\,,
\end{eqnarray}
where $\gamma$ is replaced by $-\gamma_0$. This is done to preserve the bound condition of the solution, \textit{i.e.}, $\phi(t(x_2)\rightarrow \infty)=0$ and $f(t)$ is expected to behave like $f(t(x_2)\rightarrow 0)=0$. Under this circumstances we have $\widetilde{\lambda}_0=-\gamma_0^2$. Now introducing $\chi(y)=\mathcal{L}\left\{{f(t)}\right\}$ in Eq. (22), and using Eq. (3), and Eq. (4), we have a first order differential equation in transformed space as
\begin{eqnarray}
\left\{(y^2-\gamma_2^2)\frac{d}{dy}+(2\gamma_{0}+1)y+\gamma_1^2\right\}\chi(y)=0\,,
\end{eqnarray}
which has a solution
\begin{eqnarray}
\chi(y)=K_1(y+\gamma_2)^{-(2\gamma_{0}+1)}\Big(\frac{y-\gamma_2}{y+\gamma_2}\Big)^{-\frac{\gamma_1^2}{2\gamma_2}-\frac{2\gamma_{0}+1}{2}}\,,
\end{eqnarray}
where $K_1$ is the integral constant. The last multiplier in Eq. (24) is a multivalued function when the power is a non integer. As the wave function must be single valued, we must take
\begin{eqnarray}
-\frac{\gamma_1^2}{2\gamma_2}-\frac{2\gamma_{0}+1}{2}=n\,\, (n=0, 1, 2, \ldots)\,,
\end{eqnarray}
which determines the bound states as a quantization rule for the system. Following the above quantization condition along with Eq. (14), and the relations $\widetilde{\lambda}_i=\frac{\lambda_i}{\alpha^2}=-\gamma_i^2$ give the energy eigenvalues as
\begin{eqnarray}
E_n=-\kappa\Lambda\pm\sqrt{(\kappa^2-1)[\Lambda^2-(p_1^2+p_3^2+m^2)]}\,,
\end{eqnarray}
where $\kappa=\frac{\beta_0}{\beta_1}$ and $\Lambda=\alpha(n+\frac{1}{2})\sqrt{1-\kappa^2}-p_1$.

Eq. (24) with the help of Eq.(25) gives $\chi(y)$ as
\begin{eqnarray}
\chi(y)=K_1(y+\gamma_2)^{-a}(y-\gamma_2)^{-b}=K_1g(y)h(y)\,,
\end{eqnarray}
where $a=(2\gamma_{0}+n+1)$ and $b=-n$, and the functions $g(y)$, and $h(y)$ help us for using the convolution theorem.
The original function $f(t)$ can be extracted from the convolution theorem given by Eq. (7). Following the Refs. [28-30], we can achieve
\begin{eqnarray}
f(t)=\mathcal{L}^{-1}\left\{\chi(y)\right\}=K_1\frac{e^{-\gamma_2 t}}{\Gamma(a+b)}t^{a+b-1}\,_{1}F_{1}(b; a+b; 2\gamma_2 t)\,,
\end{eqnarray}
and finally we obtain the whole eigenfunctions as
\begin{eqnarray}
\phi(x_2)=t^{-\gamma_{0}}f(t)=K_1\frac{e^{-\gamma_2 e^{-\alpha x_2}}}{\Gamma(2\gamma_{0}+1)}e^{-\gamma_0\alpha x_2}\,_{1}F_{1}(-n; 2\gamma_{0}+1; 2\gamma_2 e^{-\alpha x_2})\,.
\end{eqnarray}
where $\,_{1}F_{1}(a';b';x)$ is the confluent hypergeometric functions of the first kind [35]. We see that these above results are consistent with the previous work listed in Ref. [17].

\subsection{Case-II}
In this case, we know that $t=\alpha x_2$ and $s(x_2)=t^{-1}=\frac{1}{\alpha x_2}$. This gives the four vector potential $\{A_{0},\vec{\bold{A}}\}=\left\{\frac{\beta_0^{'}}{x_2},\frac{\beta_1^{'}}{x_2},0,0\right\}$ which generates electric and magnetic fields as $\vec{E}=\frac{\beta_0^{'}}{x_2^2}\hat{x}_2$, and $\vec{B}=\frac{\beta_1^{'}}{x_2^2}\hat{x}_3$, respectively with $\beta_0^{'}=\frac{\beta_0}{\alpha}$ and $\beta_1^{'}=\frac{\beta_1}{\alpha}$. This configuration of vector potential gives inverse-squared external fields that are orthogonal to each other.

As previous to get a bound state eigenfunctions we take $\gamma=-\gamma_2$ . Hence $\widetilde{\lambda}_2=-\gamma_2(\gamma_2+1)$ as $m_1=0$. Further, by setting $\widetilde{\lambda}_1=-\gamma_1^2$ and $\widetilde{\lambda}_0=-\gamma_0^2$ Eq. (19) provides
\begin{eqnarray}
\left[t\frac{d^2}{dt^2}-2\gamma_2\frac{d}{dt}-\gamma_1^2-\gamma_0^2 t\right]f(t)=0\,,
\end{eqnarray}
Introducing $\chi(y)=\mathcal{L}\left\{{f(t)}\right\}$ along with Eq. (3) and Eq. (4) we have
\begin{eqnarray}
\left\{(y^2-\gamma_0^2)\frac{d}{dy}+2(\gamma_2+1)y+\gamma_1^2\right\}\chi(y)=0\,,
\end{eqnarray}
which has the solution
\begin{eqnarray}
\chi(y)=K_2(y+\gamma_0)^{-2(\gamma_2+1)}\Big(\frac{y-\gamma_0}{y+\gamma_0}\Big)^{-\frac{\gamma_1^2}{2\gamma_0}-(\gamma_2+1)}\,,
\end{eqnarray}
where $K_2$ is the integral constant. As the eigenfunctions must be single valued, we must take
\begin{eqnarray}
-\frac{\gamma_1^2}{2\gamma_0}-(\gamma_2+1)=n\,\, (n=0, 1, 2, \ldots)\,,
\end{eqnarray}
For a physically accepted solution $\gamma_2$ should be $-\frac{1}{2}+\sqrt{\frac{1}{4}-\widetilde{\lambda}_2}$ where $\widetilde{\lambda}_2=\frac{\lambda_2}{\alpha^2}$. Such choice preservers the bound type solution, \textit{i.e.}, $\phi(t(x_2)\rightarrow \infty)=0$ and $f(t)$ is expected to behave like $f(t(x_2)\rightarrow 0)=0$.

Now following the quantization condition Eq. (33) along with Eq. (14) via the relations $\widetilde{\lambda}_i=\frac{\lambda_i}{\alpha^2}=-\gamma_i^2\,\, (i=0,1)$ the energy eigenvalue emerges as
\begin{eqnarray}
E_n=\frac{1}{q^2\kappa^2\beta_1^{'2}+\Lambda^2}\Big[\kappa\beta_1^{'2}p_1q^2\pm\Lambda\sqrt{M^2(\Lambda^2+q^2\kappa^2\beta_1^{'2})-q^2\beta_1^{'2}p_1^2}\Big]\,,
\end{eqnarray}
where $\Lambda=(n+\frac{1}{2})+\sqrt{\frac{1}{4}+q^2\beta_1^{'2}(1-\kappa^2)}$ , $\kappa=\frac{\beta_1^{'2}}{\beta_1^{'2}}$ and $M^2=p_1^2+p_3^2+m^2$.

Using Eq. (33), Eq. (32) can be written as
\begin{eqnarray}
\chi(y)=K_2(y+\gamma_0)^{-a}(y-\gamma_0)^{-b}=K_2g(y)h(y)\,,
\end{eqnarray}
where $a=2(\gamma_2+1)+n$ and $b=-n$.
Following the Refs. [28-30] by the means of convolution theorem as previous, we can achieve
\begin{eqnarray}
f(t)=\mathcal{L}^{-1}\left\{\chi(y)\right\}=K_2\frac{e^{-\gamma_0 t}}{\Gamma(a+b)}t^{a+b-1}\,_{1}F_{1}(b;a+b;2\gamma_0 t)\,,
\end{eqnarray}
and
\begin{eqnarray}
\phi(x_2)=K_2\frac{e^{-\alpha\gamma_0 x_2}}{\Gamma(2\gamma_2+2)}(\alpha x_2)^{\gamma_2+1}\,_{1}F_{1}(-n;2(\gamma_2+1);2\alpha\gamma_0  x_2)\,.
\end{eqnarray}
with the confluent hypergeometric functions of the first kind $\,_{1}F_{1}(a';b';z)$ [35]. We observe that these results are consistent with the work [16] when $\beta_0^{'}\equiv\kappa=0$. It is worth to mention here that there exist an interesting symmetry transformation between Case-I and Case-II. The substitution $\gamma_2\rightarrow \gamma_0$ and $\gamma_0\rightarrow \gamma_2+\frac{1}{2}$ in Eq. (22) generates Eq. (30). Physically this is important for the realization of the behavior of spin-$0$ particles when it passes from the inverse square law type field to the exponential decay type field.

\subsection{Case-III}
\subsubsection{III(a)}
In this case, we have $\widetilde{\lambda}_1=-\gamma(\gamma+1)$ and $t=\frac{1}{\alpha x_2}$. The vector potential comes out $\{A_{0},\vec{\bold{A}}\}=\left\{\frac{\beta_0^{'}}{x_2^2},\frac{\beta_1^{'}}{x_2^2},0,0\right\}$ where $\beta_i^{'}=\frac{\beta_i}{\alpha^2}$. This generates the field profile as $\vec{E}=\frac{2\beta_0^{'}}{x_2^3}\hat{x}_2$ and $\vec{B}=\frac{2\beta_1^{'}}{x_2^3}\hat{x}_3$ which are orthogonal, inverse-cubic fields.

As previous applying the Laplace  transformation in Eq. (20a) by means of the notation $\chi(y)=\mathcal{L}\left\{{f(t)}\right\}$ we have
\begin{eqnarray}
\chi(y)=K_{3a}(y^2+\widetilde{\lambda}_{2})^{\gamma}\,,
\end{eqnarray}
where $K_{3a}$ is the integration constant.

To find the solution in real space we have to take inverse transform of Eq. (38) and this is possible if we choose $-\gamma=n\,\,(n=1, 2, 3, \ldots)$. The following formula is useful here [35]
\begin{eqnarray}
\mathcal{L}^{-1}\left\{{\frac{1}{(y^2+a^2)^l}}\right\}=\frac{\sqrt{\pi}}{\Gamma (l)}\Big(\frac{t}{2a}\Big)^{l-\frac{1}{2}}J_{l-\frac{1}{2}}(at)\,, \,\, l>0\,,
\end{eqnarray}
where $J_{l-\frac{1}{2}}(at)$ is the Bessel function of the first kind. Hence we obtain the solution of Eq. (19a) as $f(t)=K_{3a}\frac{\sqrt{\pi}}{\Gamma (n)}\left(\frac{t}{2\frac{\sqrt{\lambda_2}}{\alpha}}\right)^{n-\frac{1}{2}}J_{n-\frac{1}{2}}\Big(\frac{\sqrt{\lambda_2}}{\alpha}t\Big)$.
Finally we have the eigenfunctions for the situation as
\begin{eqnarray}
\phi(x_2)\approx x_2^{1/2}J_{n-\frac{1}{2}}(\frac{\sqrt{\lambda_2}}{\alpha^2 x_2})\,,
\end{eqnarray}
with the energy eigenvalues
\begin{eqnarray}
E_n=\frac{1}{2q\beta_0{'}}[2q\beta_1^{'}p_1+n(n-1)]\,.
\end{eqnarray}

\subsubsection{III(b)}
As previous here we have $\widetilde{\lambda}_1=-\gamma(\gamma+\frac{1}{2})$ and $t=\frac{4}{\alpha^2 x_2^2}$ with $m_1=3/2$ hence $m_2=2(m_1-1)=1$. The four vector potential $\{A_{0},\vec{\bold{A}}\}=\left\{\frac{\beta_0^{'}}{x_2^2},\frac{\beta_1^{'}}{x_2^2},0,0\right\}$ where $\beta_i^{'}=\frac{4\beta_i}{\alpha^2}$. This generates the field profile as $\vec{E}=\frac{2\beta_0^{'}}{x_2^3}\hat{x}_2$ and $\vec{B}=\frac{2\beta_1^{'}}{x_2^3}\hat{x}_3$. Applying the Laplace  transformation in Eq. (20b) by means of the notation $\chi(y)=\mathcal{L}\left\{{f(t)}\right\}$ we have
\begin{eqnarray}
\chi(y)=K_{3b}y^{-2\gamma_1-\frac{1}{2}}e^{-\frac{\widetilde{\lambda}_{2}}{y}}\,,
\end{eqnarray}
where we have used $\gamma=-\gamma_1$. To get the inverse Laplace transform we select $2\gamma_1+1/2=n\,\,(n=1,2,3\ldots)$. The formula [35] is useful in this purpose
\begin{eqnarray}
\mathcal{L}^{-1}\left\{{\frac{1}{y^\mu}}e^{-\frac{k}{y}}\right\}=\Big(\frac{t}{k}\Big)^{\frac{\mu-1}{2}}J_{\mu-1}(2\sqrt{kt})\,,\,\ \mu>0\,.
\end{eqnarray}
So we have the solution as $f(t)=K_{3b}\Big(\frac{t}{\widetilde{\lambda}_2}\Big)^{\frac{n-1}{2}}J_{n-1}\Big(2\sqrt{\widetilde{\lambda}_{2}t}\Big)$.
The eigenfunctions emerge as
\begin{eqnarray}
\phi(x_2)\approx x_2^{1/2}J_{n-1}(\frac{4\sqrt{\lambda_2}}{\alpha^2x_2})\,,
\end{eqnarray}
with the energy eigenvalues
\begin{eqnarray}
E_n=\frac{1}{2q\beta_0^{'}}\Big[2q\beta_1^{'}p_1+(n-1)^2-1/4\Big]\,.
\end{eqnarray}
Manipulating the parameters in section III originates Eq. (20a) and (20b). Study of Case-III(a) and (b) provides same type of external fields with different magnitudes. Also the energy eigenvalues for these cases are almost same with slight difference, which is clearly seen from Eqs. (41) and (45).

\subsection{Case-IV}
\subsubsection{IV(a)}
Here the Laplace transform of Eq. (21a) provides
\begin{eqnarray}
\chi(y)=K_{4a}(y^2+\widetilde{\lambda}_0)^{-\gamma_1-1}\,,
\end{eqnarray}
where $\gamma=-\gamma_1$ and $K_{4a}$ is the integration constant. Hence we have the field profile as $\{A_{0},\vec{\bold{A}}\}=\left\{\frac{\beta_0^{'}}{x_2^2},\frac{\beta_1^{'}}{x_2^2},0,0\right\}$ where $\beta_i^{'}=\frac{\beta_i}{\alpha^2} (i=0,1)$. This will facilitate the electromagnetic field as $\vec{E}=\frac{2\beta_0^{'}}{x_2^3}\hat{x}_2$ and $\vec{B}=\frac{2\beta_1^{'}}{x_2^3}\hat{x}_3$. To get an inverse Laplace transformation of Eq. (46) we must use the relation given by Eq. (39) as earlier. After setting quantization condition $\gamma_1+1=n\,\,(n=1,2,3\ldots)$, we have $f(t)=K_{4a}\frac{\sqrt{\pi}}{\Gamma(n)}\Big(\frac{t}{2\frac{\sqrt{\lambda_0}}{\alpha}}\Big)^{n-\frac{1}{2}}J_{n-\frac{1}{2}}\Big(\frac{\sqrt{\lambda_0}}{\alpha}t\Big)$. The eigenfunctions for this case is
\begin{eqnarray}
\phi(x_2)\approx x_2^{1/2}J_{n-\frac{1}{2}}(\sqrt{\lambda_0}x_2)\,,
\end{eqnarray}
with energy eigenvalues
\begin{eqnarray}
E_n=\frac{1}{2q\beta_0^{'}}\Big[2qp_1\beta_1^{'}+n(n-1)\Big]\,.
\end{eqnarray}
\subsubsection{IV(b)}
The Laplace transform of Eq. (21b) with the setting $\gamma=-\gamma_1$ provides
\begin{eqnarray}
\chi(y)=K_{4b} y^{-2\gamma_1-\frac{3}{2}}e^{-\frac{\widetilde{\lambda}_0}{y}}\,,
\end{eqnarray}
The field profile in this case is $\{A_{0},\vec{\bold{A}}\}=\left\{\frac{\beta_0^{'}}{x_2^2},\frac{\beta_1^{'}}{x_2^2},0,0\right\}$ where $\beta_i^{'}=\frac{4\beta_i}{\alpha^2}\,\,(i=0,1)$. Using the relation given by Eq. (43) as previous we have the solution $f(t)=K_{4b}\Big(\frac{t}{\widetilde{\lambda}_0}\Big)^{\frac{n-1}{2}}J_{n-1}(2\sqrt{\widetilde{\lambda}_{0}t})$ which gives
\begin{eqnarray}
\phi(x_2)\approx x_2^{1/2}J_{n-1}(\sqrt{\lambda_0} x_2)\,,
\end{eqnarray}
where the quantization condition is used as $2\gamma_1+\frac{3}{2}=n\,\,(n=1,2,3\ldots)$.
The energy eigenvalues for this case will be
\begin{eqnarray}
E_n=\frac{1}{2q\beta_0^{'}}\Big[2qp_1\beta_1^{'}+(n-1)^2-\frac{1}{4}\Big]\,.
\end{eqnarray}

At this point, we underline that Eq. (20a) for Case-III and Eq. (21a) for Case-IV are symmetric under the replacements $\gamma \rightarrow \gamma-1$ and $\widetilde{\lambda}_2 \rightarrow \widetilde{\lambda}_0$ while Eq. (20b) and Eq. (21b) are symmetric under the replacements $\gamma \rightarrow \gamma-1/2$ and $\widetilde{\lambda}_2 \rightarrow \widetilde{\lambda}_0$. Thus electromagnetic field profiles are same for these cases. So, we expect that a spin-$0$ particle has the same energy subjected to the same em fields which are obtained for different mathematical settings. However, the eigenfunctions written for the above situations have a minor difference as $\widetilde{\lambda}_2 \rightarrow \widetilde{\lambda}_0$ which does not effect the energy eigenvalues because the parameter $\widetilde{\lambda}_2$ is independent from the energy.

\section{Conclusion}

We have obtained the exact solutions of the KG equation for three different electromagnetic field configurations, namely exponential, inverse square and inverse cubic law type fields, in closed forms. For this aim, we have generated these fields from the study of a special type of differential equation obtained from KG equation. We have used the Laplace transform for finding the solutions for bound states and seen that this formalism is suitable for analyzing the motion of a charged particle in external em fields. The generic differential equation has been transformed to a second order one with coefficients linear in the independent variable by the means of four different options giving distinct mathematical settings for the parameters. Out of these four options, we have obtained three different external field configurations and studied them in details. For the first two cases, where electric and magnetic fields are exponential and inverse squared type, we have computed the spectrum of spin-$0$ particles and they are comparable with the previous works carried out in the literature. The last two cases are for inverse cubic type of field. Obviously the energy eigenvalues equations for these cases are identical but the expression of $\phi(x_2)$ changes significantly. In Case-III  where the Bessel function's argument contains the space variable $x_2$ as inverse whereas the same space variable comes in a linear way (in the argument of Bessel function) for the Case-IV.\\
Also we have extracted that, there exists a symmetry between Case-I and Case-II. The substitution $\gamma_2\rightarrow \gamma_0$ and $\gamma_0\rightarrow \gamma_2+\frac{1}{2}$ in Eq. (22) generates Eq. (30). Similarly, Eq. (20a) for Case-III and Eq. (21a) for Case-IV are symmetric under the replacements $\gamma \rightarrow \gamma-1$ and $\widetilde{\lambda}_2 \rightarrow \widetilde{\lambda}_0$ while Eq. (20b) and Eq. (21b) are symmetric under the replacements $\gamma \rightarrow \gamma-1/2$ and $\widetilde{\lambda}_2 \rightarrow \widetilde{\lambda}_0$. This type of symmetry is important for future research as they will surely explore the behavior of a relativistic spin-$0$ charged particles in space varying em fields. We expect impending works to reveal the physics behind it.

\section{Acknowledgments}
We thank the referee for suggestions which gave us deeper insights about improving of the present problem. One of authors (A.A.) thanks Prof Dr Andreas Fring from City University London and the Department of Mathematics for their hospitality where this work has began. This research was partially supported by through a fund provided by University of Hacettepe.

\end{document}